\documentclass[pdflatex,default,iicol,Numbered]{sn-jnl}

\usepackage{lineno}

\usepackage{graphicx}%
\usepackage{multirow}%
\usepackage{amsmath,amssymb,amsfonts}%
\usepackage{amsthm}%
\usepackage{mathrsfs}%
\usepackage[title]{appendix}%
\usepackage{xcolor}%
\usepackage{textcomp}%
\usepackage{manyfoot}%
\usepackage{booktabs}%
\usepackage{algorithm}%
\usepackage{algorithmicx}%
\usepackage{algpseudocode}%
\usepackage{listings}%

\usepackage{arydshln}
\usepackage{pgf,tikz}
\usetikzlibrary{shapes,backgrounds,arrows,calc,automata,tikzmark,positioning}

\definecolor{light-gray}{gray}{0.5}

\begin{document}

\title[CPTR]{Constrained Pressure-Temperature Residual (CPTR) Preconditioner Performance for Large-Scale Thermal CO$_2$ Injection Simulation}

\author*[1]{\fnm{Matthias A.} \sur{Cremon}}\email{mcremon@stanford.edu}

\author[1]{\fnm{Jacques} \sur{Franc}}\email{jfranc@stanford.edu}

\author[2]{\fnm{Fran\c cois P.} \sur{Hamon}}\email{francois.hamon@totalenergies.com}

\affil*[1]{\orgdiv{Energy Science Engineering}, \orgname{Stanford University}, \orgaddress{\city{Stanford}, \postcode{94305}, \state{CA}, \country{USA}}}

\affil[2]{\orgname{TotalEnergies E\&P Research \& Technology}, \orgaddress{\city{Houston}, \postcode{77002}, \state{TX}, \country{USA}}}










\maketitle

\begin{abstract}

This work studies the performance of a novel preconditioner, designed for thermal reservoir simulation cases and recently introduced in \citet{Roy20} and \citet{Cremon20etal}, on large-scale thermal CO$_2$ injection cases. For Carbon Capture and Sequestration (CCS) projects, injecting CO$_2$ under supercritical conditions is typically tens of degrees colder than the reservoir temperature. Thermal effects can have a significant impact on the simulation results, but they also add many challenges for the solvers. More specifically, the usual combination of an iterative linear solver (such as GMRES) and the Constrained Pressure Residual (CPR) physics-based block-preconditioner is known to perform rather poorly or fail to converge when thermal effects play a significant role. The Constrained Pressure-Temperature Residual (CPTR) preconditioner retains the $2\times2$ block structure (elliptic/hyperbolic) of CPR but includes the temperature in the elliptic subsystem. Doing so allows the solver to appropriately handle the long-range, elliptic part of the parabolic energy equation. The elliptic subsystem is now formed by two equations, and is dealt with by the system-solver of BoomerAMG (from the HYPRE library). Then a global smoother, ILU(0), is applied to the full system to handle the local, hyperbolic temperature fronts.
We implemented CPTR in the multi-physics solver GEOS and present results on various large-scale thermal CCS simulation cases, including both Cartesian and fully unstructured meshes, up to tens of millions of degrees of freedom. The CPTR preconditioner severely reduces the number of GMRES iterations and the runtime, with cases timing out in 24h with CPR now requiring a few hours with CPTR. We present strong scaling results using hundreds of CPU cores for multiple cases, and show close to linear scaling. CPTR is also virtually insensitive to the thermal P\'eclet number (which compares advection and diffusion effects) and is suitable to any thermal regime.

\end{abstract}



\keywords{Carbon Capture and Sequestration, Thermal CO$_2$ Injection, Multi-stage Preconditioning, Iterative Methods}




\section{Introduction}
\label{sec:intro}

Carbon Capture and Sequestration (CCS) is one of the leading technologies to reduce carbon emissions in the near future \citep{kelemen2019overview,ipcc_2022}, and has been extensively studied both in experimental and numerical studies in recent years \citep{Pruess04, Class09, Eigestad2009, Nordbotten12, Flemish23, spe11}. CCS cases consider three types of reservoirs: saline aquifers, depleted hydrocarbons reservoirs \citep{de2014carbon} and mafic formations \citep{gunnarsson2018rapid}. Overall, existing storage site candidates (saline and depleted reservoirs) are estimated to be able to store between 5,000 and 25,000 Gt CO$_2$ over hundreds of years. The governing equations and general numerical simulation features of CCS are virtually identical to reservoir simulation processes used in the oil and gas industry for decades \citep{Aziz79}, and many of the numerical simulators capable of running reservoir simulations have been updated to handle CO$_2$ injection \citep{eclipse, adgprs, intersect}. Some of the main features of CCS simulations differ from oil production simulations, including but not limited to: the injection schemes and the absence of producers, the importance of geomechanical and geochemichal couplings, the fluid phase behavior and properties, and some of the most relevant physical processes.

Importantly for this work, the dependency of CCS processes on thermal effects is well known \citep{Han10}. The fluid properties largely vary when CO$_2$ is injected tens of degrees colder than the reservoir temperature. For example, the Joule-Thomson (JT) effect plays an important role for CO$_2$ injection \citep{Oldenburg07, Pruess11, Vilarrasa17, Wapperom22}. When the JT effect is modeled, the energy equation needs to be added to the set of global equations, since a temperature drop occurs and the processes are not isothermal anymore, even if the injection and reservoir temperatures are the same. In CCS, due to the small number of injectors and the size of the reservoirs, the long-range elliptic-like parts of the energy equation likely dominate the energy transport outside of a very small near-well region, resulting in a challenging situation for the numerical solvers.

Physics-based block-preconditioners are among the most active research areas in reservoir simulation, with a focus on multi-physics cases involving flow and geomechanics \citep{ThcJia09, White11, Voskov12b, Haga12, Zhou13, Gries14, Cusini15, White16, Gaspar17, White19}. For flow and transport problems (i.e., assuming no deformation of the rock matrix), most models use the two-stage Constrained Pressure Residual (CPR) preconditioner \citep{Wallis83,Wallis85}. In an isothermal simulation, only one equation exhibits an elliptic behavior. CPR leverages that and extracts that equation into a so-called pressure system, which will be approximated by a first-stage preconditioner well suited for elliptic systems. Then the second stage is a generic local smoother that can address local errors. In the case of thermal simulations, CPR has been known to show convergence issues primarily due to the parabolic nature of the energy equation. Conduction of heat through both the rock matrix and the phases cannot be neglected \citep{Incropera07}, and the lack of appropriate treatment for the elliptic part of the energy equation equation in the standard CPR algorithm renders the method inadequate \citep{Li15,Li17}.

In this paper, we implemented and tested an extended version of the CPR preconditioner introduced in \citet{Roy20} and \citet{Cremon20etal}, named Constrained Pressure-Temperature Residual (CPTR). As the name suggests, both the pressure equation and the energy equation are extracted in the first stage, allowing the preconditioner to properly address the elliptic part of the energy equation. To our knowledge, this is the first implementation and in-depth study of the performance of the CPTR preconditioner in a classical, finite-volume based reservoir simulation software with CPU parallelism. The implementation can be used for both multiphase hydrocarbon recovery scenarios and carbon sequestration cases, and we focus on the latter in this work. We test the algorithm on multiple large-scale cases based on real reservoir models, with sizes spanning from hundreds to tens of millions of degrees of freedom. Those cases are comparable in size to recently published hydrocarbon-based \citep{cao2021adding,tene2023graphics}, geothermal \citep{khait2020high} and CCS \citep{panfili2022, kachuma2023assessment} simulation studies using real reservoir models (i.e. not synthetic datasets) and HPC hardware, although none of them consider thermal effects. Most of the test cases are real candidates for CO$_2$ injection in the near future, and show that the algorithm performs extremely well and outperforms CPR by an order of magnitude. We also verify that the solutions using CPTR scale linearly using hundreds of CPU cores, and that the sensitivity to the thermal regime, as described by the thermal P\'eclet number, is negligible.

The paper is structured as follows: Section \ref{sec:goveq} briefly describes the governing equations, Section \ref{sec:precs} shows the several block-preconditioners considered in this work, Section \ref{sec:cases} discusses the various test cases used to evaluate the preconditioners performance, Section \ref{sec:results} shows all of our numerical results, including strong scaling and sensitivity to the thermal P\'eclet number, and finally, Section \ref{sec:conclusion} gives a conclusion and discusses future work avenues.

\section{Problem Statement}
\label{sec:goveq}

We consider a general system of $n_c$ components and $n_p$ phases, with compositional and thermal descriptions, resulting in a coupled system of conservation equations. The mass conservation equations read
\begin{align}
\begin{split}\label{eq::mass}
 &\dfrac{\partial}{\partial t}\left(\phi\sum\limits_{p=1}^{n_p}x_{cp}\rho_pS_p\right) + \nabla \cdot \left(\sum\limits_{p=1}^{n_p}x_{cp}\rho_p\textbf{u}_p\right) \\ + & \sum\limits_{p=1}^{n_p} x_{cp}\rho_pq_p = 0, \hspace{0.6cm} c = 1,\dots,n_c \,,
\end{split}
\end{align}

where $\phi$ is the porosity; $x_{cp}$ is the mole fraction of component $c$ in phase $p$; $\rho_p$, $S_p$, $\textbf{u}_p$, and $q_p$ are the molar density, saturation, velocity, and volumetric flow rate of phase $p$, respectively. In the remainder of this work, we have two components, CO$_2$ and water, split in two phases (i.e. $n_c = n_p = 2$).

The energy conservation equation reads
\begin{align}
\begin{split} \label{eq::energy}
& \dfrac{\partial}{\partial t}\left(\phi\sum\limits_{p=1}^{n_p}U_{p}\rho_pS_p+(1-\phi) \tilde U_r\right) \\ + & \nabla \cdot \left(\sum\limits_{p=1}^{n_p}H_{p}\rho_p\textbf{u}_p\right) - \nabla \cdot \left(\kappa\nabla T\right) \\ + & \sum\limits_{p=1}^{n_p} H_p\rho_pq_p = 0\,,
\end{split}
\end{align}
where $T$ is the temperature; $\tilde U_r$ is the rock volumetric internal energy; $\kappa$ is the thermal conductivity; $U_{p}$ and $H_{p}$ are the internal energy and enthalpy of phase $p$, respectively.

This system of equations is solved using the GEOS \citep{geosx} multiphysics solver, jointly developed between Lawrence Livermore National Laboratory, Stanford University, TotalEnergies, and Chevron. The fluid models employed to perform component mass partitioning between phases (flash) and to compute fluid properties as a function of pressure, temperature, and component fractions are listed in the GEOS documentation.

\section{Preconditioners}
\label{sec:precs}

The preconditioners considered in this work are all physics-based, two-stage block preconditioners. The general form of a two-stage preconditioner is
\begin{equation} \label{eq::multistageprec}
M^{-1} = M_1^{-1} + M_2^{-1} \left(I - AM_1^{-1}\right) \,,
\end{equation}
with $M_1^{-1}$ and $M_2^{-1}$ the first- and second-stage preconditioners, respectively, and $A$ the Jacobian matrix. $A$ is decomposed in a $2\times2$ block matrix as follows
\begin{equation*}
A = \begin{bmatrix} A_{hh} & A_{he} \\ A_{eh} & A_{ee} \end{bmatrix},
\end{equation*}
with $e$ and $h$ the elliptic and hyperbolic variables, respectively. The preconditioners in this work differ in the way our four primary variables (two global component densities, pressure and temperature) are assigned to the elliptic or hyperbolic blocks. The first-stage preconditioner has the form
\begin{equation*} \label{eq::CPRM1}
M_1^{-1} = \begin{bmatrix} 0 & 0 \\ 0 & M_{ee}^{-1} \end{bmatrix},
\end{equation*}
and the second-stage preconditioner is applied to the global matrix. Both stages use preconditioners from the HYPRE libary \citep{Falgout02} and are paired with a right-preconditioned GMRES \citep{Saad86} linear solver, with 200 maximum iterations and a relative tolerance of $10^{-6}$.  All Schur complement and block matrix operations are handled by the MultiGrid Reduction (MGR) framework \citep{Bui20}, also available through HYPRE. Subsections \ref{sec:cpr} and \ref{sec:cptr} describe the two preconditioners used in this work in more details, and subsection \ref{sec:mgr} discusses some implementation specifics.

\subsection{Constrained Pressure Residual (CPR)}
\label{sec:cpr}

The Constrained Pressure Residual (CPR) preconditioner was introduced in \citet{Wallis83} and \citet{Wallis85}, and leverages the fundamental difference between elliptic and hyperbolic equations. The former type results in long-range effects, while the latter shows local variations (including shocks). CPR methods have been applied to a wide range of problems, including multi-segmented wells, compositional simulations and multi-scale methods \citep{ThcJia09,Zhou13,Cusini15}. Since CPR was designed for isothermal cases, in the absence of special treatment it considers the energy equation as a secondary equation when used for thermal simulations. Multiple authors have reported convergence issues with CPR in those situations, and proposed heuristic-based solutions \citep{Li15, Li17}.

CPR is a two-stage preconditioner and relies on a block-matrix structure, with a pressure block and a secondary block. To completely describe the preconditioner, the type of method used in the first and second stage should be specified. Virtually all current implementations use an AMG-based \citep{Ruge87,Stueben01} first stage, and an ILU-based second stage, leading to CPR-AMG-ILU(k) methods \citep{Lacroix03, Cao05}. Note that all ILU methods used in this work use zero fill-in (k$\ =0$). Since all preconditioners in this work use the CPR-AMG-ILU(0) structure, we shorten the nomenclature to simply CPR. The Jacobian matrix is decomposed as
\begin{equation*}
A 
= \left[ \begin{array}{cc:c} A_{\rho\rho} & A_{\rho T} & A_{\rho p} \\ A_{T\rho} & A_{TT} & A_{Tp} \\ \hdashline A_{p\rho} & A_{pT} & A_{pp}\end{array} \right]
= \begin{bmatrix} A_{hh} & A_{hp} \\ A_{ph} & A_{pp} \end{bmatrix}.
\end{equation*}
with temperature in the hyperbolic subsystem. Note that $A_{hh}$ is shown as a $2\times2$ block matrix here for illustration purposes, but would actually have interleaved variables.
$M_{ee}^{-1}$ here uses a classical AMG preconditioner from BoomerAMG \citep{Baker11} with default parameters: one V-cycle, one aggregate level, first order relaxation, HMIS coarsening \citep{DeSterck06} and extended+i interpolation truncated to 4 elements per row \citep{DeSterck08}. The second-stage preconditioner is an ILU(0) algorithm from HYPRE \citep{Falgout02}.

\subsection{Constrained Pressure-Temperature Residual (CPTR)}
\label{sec:cptr}

The Constrained Pressure-Temperature Residual preconditioner (CPTR) is, as the name suggests, an extension of CPR and was recently introduced for multiple different test cases in \citet{Roy20} and \citet{Cremon20etal}. Due to the previously mentioned shortcomings of CPR and the parabolic nature of the energy equation, it is natural to treat temperature in the first stage to handle the long-range coupling. The Jacobian matrix is now split as
\begin{equation*}
A 
= \left[ \begin{array}{c:cc} A_{\rho\rho} & A_{\rho T} & A_{\rho p} \\ \hdashline A_{T\rho} & A_{TT} & A_{Tp} \\ A_{p\rho} & A_{pT} & A_{pp}\end{array} \right]
= \begin{bmatrix} A_{\rho\rho} & A_{\rho e} \\ A_{e\rho} & A_{ee} \end{bmatrix},
\end{equation*}
and $A_{ee}$ becomes a $2\times2$ block system. This creates some additional challenges, since it now cannot be handled by classical, monolithic AMG preconditioners. \citet{Roy19} proposed a specific block preconditioner for the $A_{ee}$ subsystem, which is tailored to a specific set of physics and discretization and cannot be generalized in a straightforward manner. In \citet{Cremon20etal}, an alternative is introduced to circumvent this problem and uses a three-stage preconditioner instead (see subsection \ref{sec:cptr3} for a brief description). The more general and natural option is to use system-AMG methods to compute $M^{-1}_{ee}$, such as SAMG \citep{Gries15} or BoomerAMG \citep{Baker11}. These packages can properly apply AMG methods to systems of equations and allow us to retain both the elliptic/hyperbolic split (and the resulting $2\times2$ block matrix) and generality, since there are no assumptions on the physics nor on the implementation. They also have shown promising early results for thermal water injection in \citet{Roy20}. Similar to the previous section, we denote the two-stage CPTR-system-AMG-ILU(0) simply as CPTR in the remainder of this work.

\subsection{Three-Stage Constrained Pressure-Temperature Residual (CPTR3)}
\label{sec:cptr3}

As we previously mentioned, in \citet{Cremon20etal} a three-stage version of the preconditioner was proposed, denoted CPTR3. The reader is referred to the initial paper for details of the algorithm, but the main idea is to simply add a third stage to the preconditioner, and use the following split for the Jacobian matrix:
\begin{equation*}
A 
= \left[ \begin{array}{c:c:c} A_{\rho\rho} & A_{\rho T} & A_{\rho p} \\ \hdashline A_{T\rho} & A_{TT} & A_{Tp} \\ \hdashline A_{p\rho} & A_{pT} & A_{pp}\end{array} \right].
\end{equation*}
Both the pressure and the temperature subsystems are extracted separately, and both are sent to a monolithic AMG preconditioner. Therefore, it bypasses the previously mentioned problem of dealing with a $2\times2$ pressure subsystem, at the cost of an extra Schur complement approximation. The structure also makes it straightfoward to add on top of an existing CPR implementation with minimal modifications. Due to the additional approximate Schur complement, CPTR3 should always perform worse than CPTR, which is also what we have observed on multiple test cases \citep{Cremon20etal}. Since our implementation in GEOS has access to the system-AMG solver of BoomerAMG, we do not consider CPTR3 in this work.

\begin{figure*}[!htb]
    \centering
    \includegraphics[width=\textwidth]{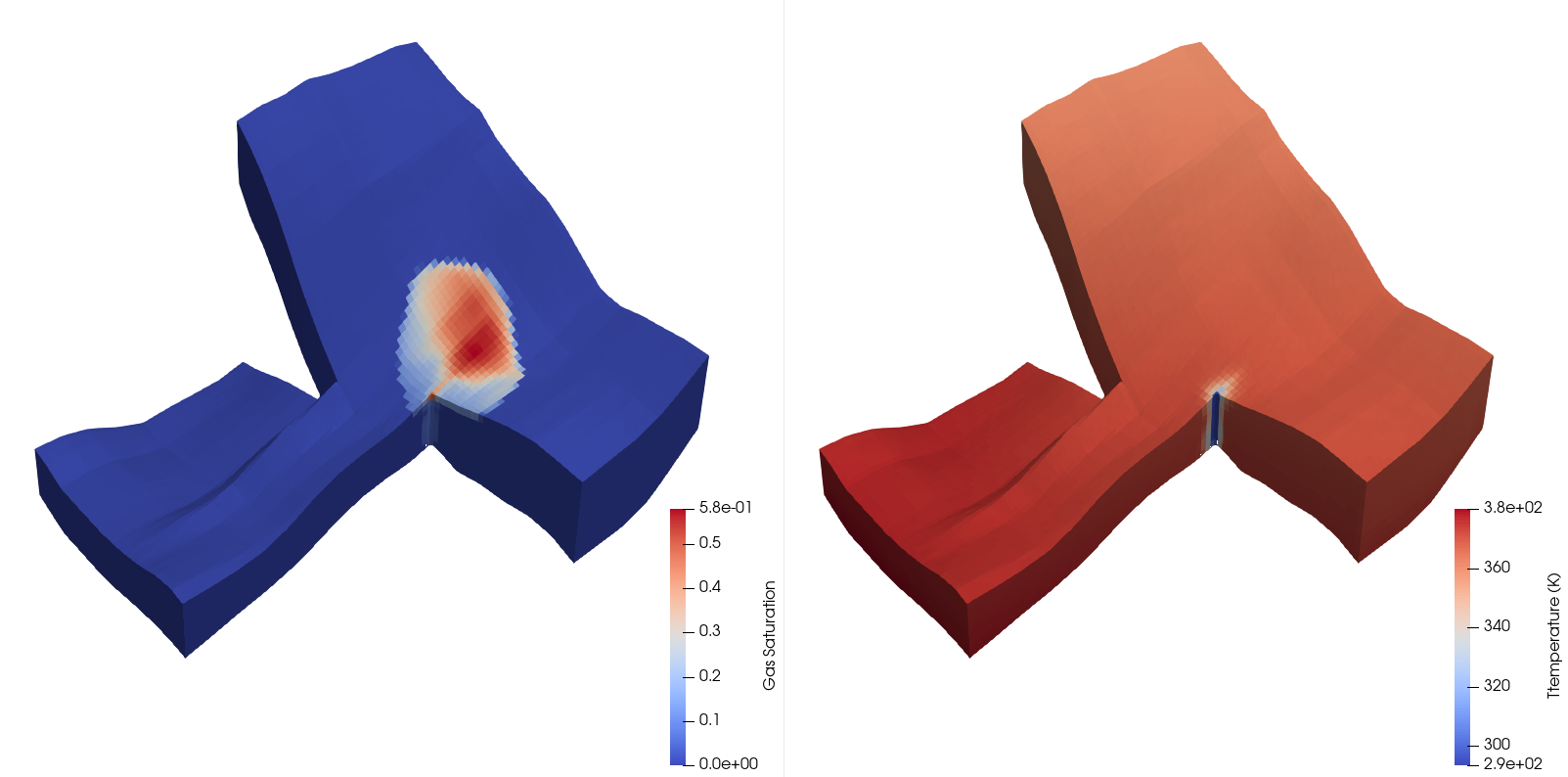}
    \caption{Gas saturation (left) and temperature (right) fields at the end of the simulation for the Johansen case. Two half cuts are used to show the well line, and the Z direction is exaggerated by a factor of 10.}
    \label{fig:johansen_field}
\end{figure*}

\subsection{A Word on Implementation}
\label{sec:mgr}

Our implementation in GEOS uses the MultiGrid Reduction (MGR \citep{Bui20}) framework from the HYPRE library \citep{Falgout02}. MGR allows multiple successive reductions to be performed to build the preconditioner, with several options to apply different smoothers at each level. For our cases, we first eliminate the last density unknown using the volume constraint equation. Since that equation is local, the Schur complement inverse is exact and we simply reduce the size of the system on which we construct a CPTR approximate inverse, with no loss of information. Then, the remaining system of $P$, $T$ and $\rho$ is used to construct a CPTR decomposition.

\section{Test Cases}
\label{sec:cases}

This work uses five different test cases, coming from four different reservoirs. All but one represent realistic saline aquifers known to be candidates for large-scale CO$_2$ sequestration projects, while the other one is a common and highly challenging benchmark case for numerical reservoir simulation. All cases are described in more details in subsections \ref{sec:johansen}, \ref{sec:spe10}, \ref{sec:hi24l}, \ref{sec:nlights}, and all the test data is summarized in subsection \ref{sec:summary}.

\subsection{Johansen}
\label{sec:johansen}

\begin{figure*}[!htb]
    \centering
    \includegraphics[width=\textwidth]{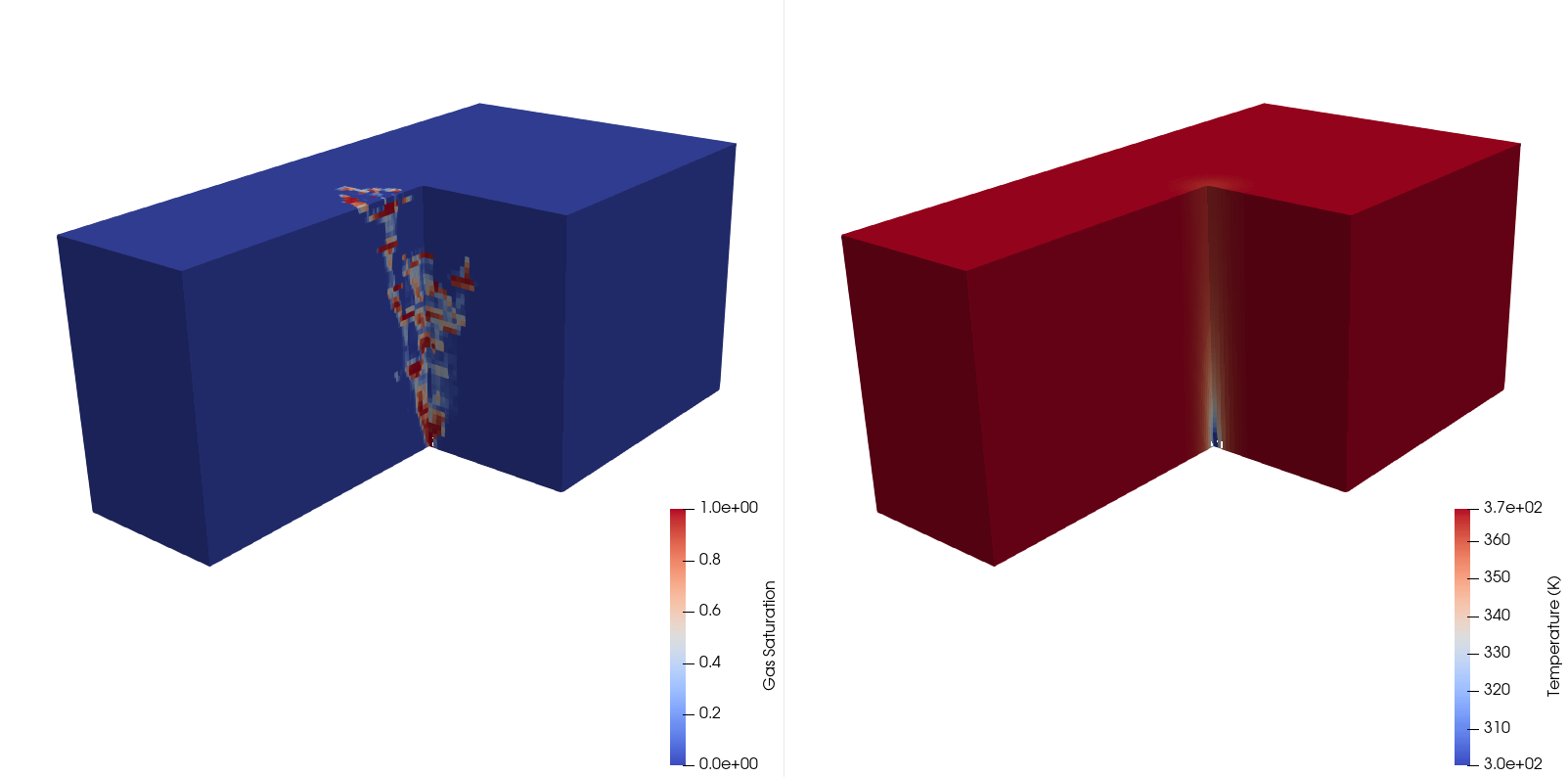}
    \caption{Gas saturation (left) and temperature (right) fields at the end of the simulation for the Modified SPE10 case. Two half cuts are used to show the well line, and the Z direction is exaggerated by a factor of 10.}
    \label{fig:SPE10_end}
\end{figure*}

We start by considering a portion of the Johansen formation, located in the Norwegian part of the North Sea, which we simply denote Johansen in the remainder of this work. The Johansen formation as a whole is a well-known candidate for CO$_2$ storage. The smaller block used in this work was initially proposed in \citet{Class09} and \citet{Eigestad2009} as a benchmark case. The reservoir is roughly $9\times10$ kilometers, with a thickness of 90-140 meters depending on the location. There is one sealed fault plane and the well is located a few hundred meters away from it. We inject pure CO$_2$ at the bottom of the formation, at a rate of 1 Mt/year for 25 years. The plume is then free to migrate for 25 more years. while a $0.03^\circ$C/m geothermal gradient is initially imposed along the reservoir depth. The injection temperature ($20^\circ$C) is about $90^\circ$C lower than the average reservoir temperature. The mesh is fully Cartesian,  uses a corner-point grid (CPG) and conforms to the fault plane, with a total of around 60,000 cells. Figure \ref{fig:johansen_field} shows the gas saturation and the temperature maps in the reservoir at the end of the simulation. The porosity of the model (the reader is referred to \citet{Class09} for details) decreases with depth, which combined with gravity effects shows the plume moving upwards in the reservoir and staying away from the fault plane.

\subsection{Modified SPE10}
\label{sec:spe10}

SPE10 is a well known dataset initially designed as the 10th comparative solution project for upscaling processes of Black-Oil simulations and was introduced in \citet{Christie01}. The initial dataset (called Model 2 in the original paper) comprises 85 layers of $220\times60$ cells. The top 35 layers represent a Tarbert formation, and the bottom 50 layers represent a channelized Upper-Ness formation. We only consider those 50 bottom layers here, for a total of 660,000 cells in the Cartesian grid. The field dimensions are significantly smaller than every other case considered in this work; the reservoir is $1,\!200\times2,\!200\times100$ feet. It is important to note that this test case is not particularly relevant as a CCS project, but due to the channelized structure and high permeability contrast, it is a highly challenging case for solvers despite its limited size. We inject from a single well in the middle of the domain, in the bottom layer, at a rate of 10 kt/year and with an injection temperature around 70$^\circ$C lower than the reservoir temperature. The only modification we made to the properties is that the vertical permeability is only 10 times smaller than the horizontal permeability (while it was 1,000 times smaller in the original dataset). This allows for more upward plume displacement, making the whole 3D pore space accessible.
We can observe on Figure \ref{fig:SPE10_end}, which shows the gas saturation and temperature in the reservoir at the end of the simulation, that the vertical connectivity is highly heterogeneous and the plume is following the channels. We refer to this test case as Modified SPE10 in the remainder of this work.

\begin{figure*}[!htb]
    \centering
    \includegraphics[width=\textwidth]{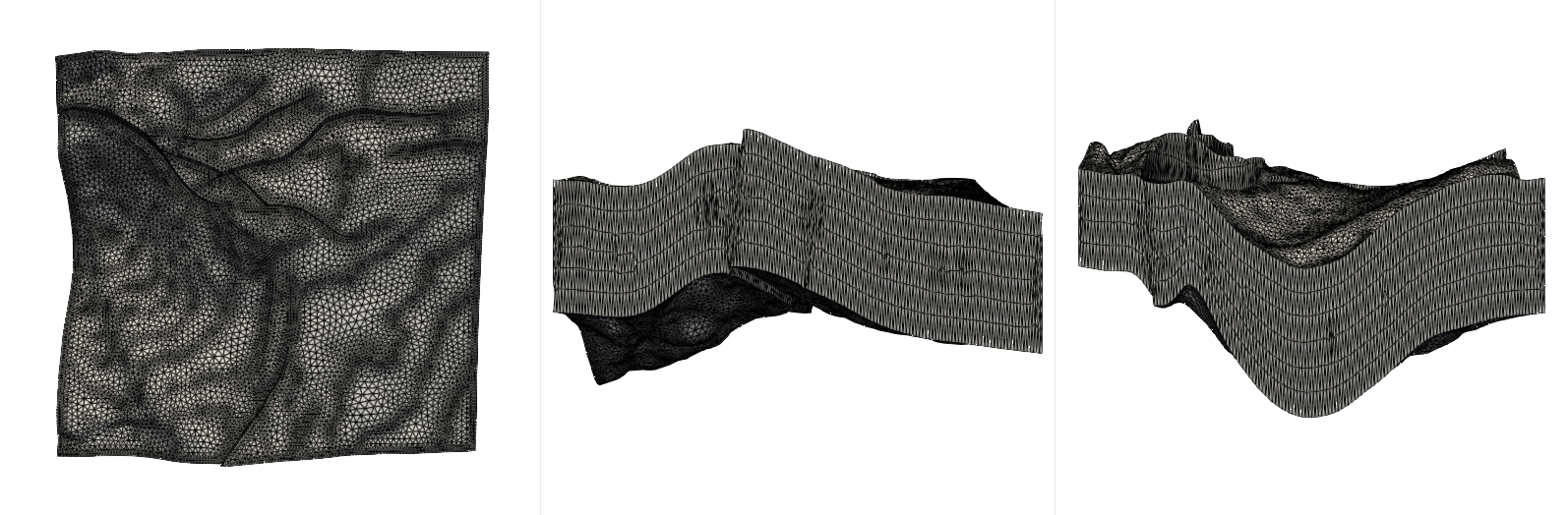}
    \caption{XY (left),  ZX (middle) and YZ (right) planes of the HI24L-S mesh. The Z direction is exaggerated by a factor of 10.}
    \label{fig:HI24L_mesh}
\end{figure*}

\begin{figure*}[!htb]
    \centering
    \includegraphics[width=\textwidth]{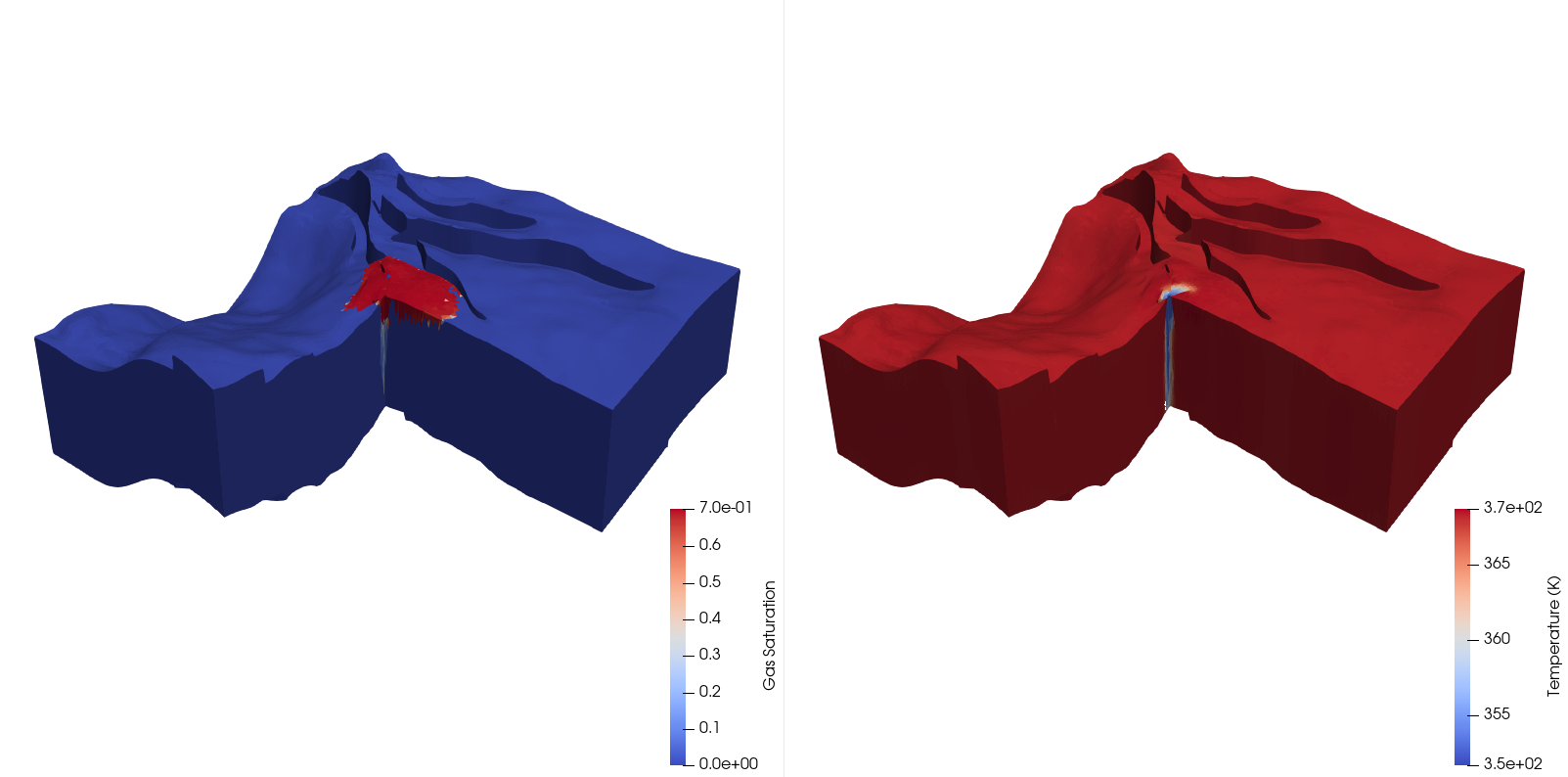}
    \caption{Gas saturation (left) and temperature (right) fields at the end of the simulation (60 years) for the HI24L-S case. Two half cuts are used to show the well line, and the Z direction is exaggerated by a factor of 10.}
    \label{fig:HI24L_end}
\end{figure*}

\begin{figure*}[!htb]
    \centering
    \includegraphics[width=\textwidth]{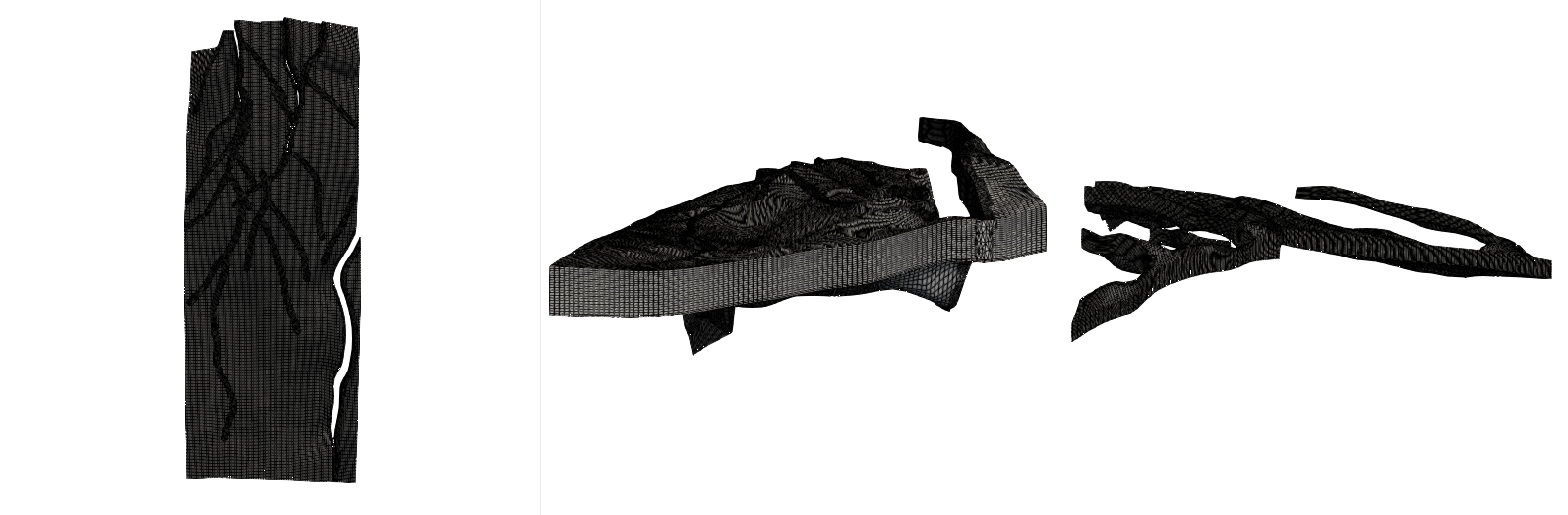}
    \caption{XY (left),  ZX (middle) and YZ (right) planes of the Northern Lights mesh. The Z direction is exaggerated by a factor of 10.}
    \label{fig:NL_mesh}
\end{figure*}

\begin{figure*}[!htb]
    \centering
    \includegraphics[width=\textwidth]{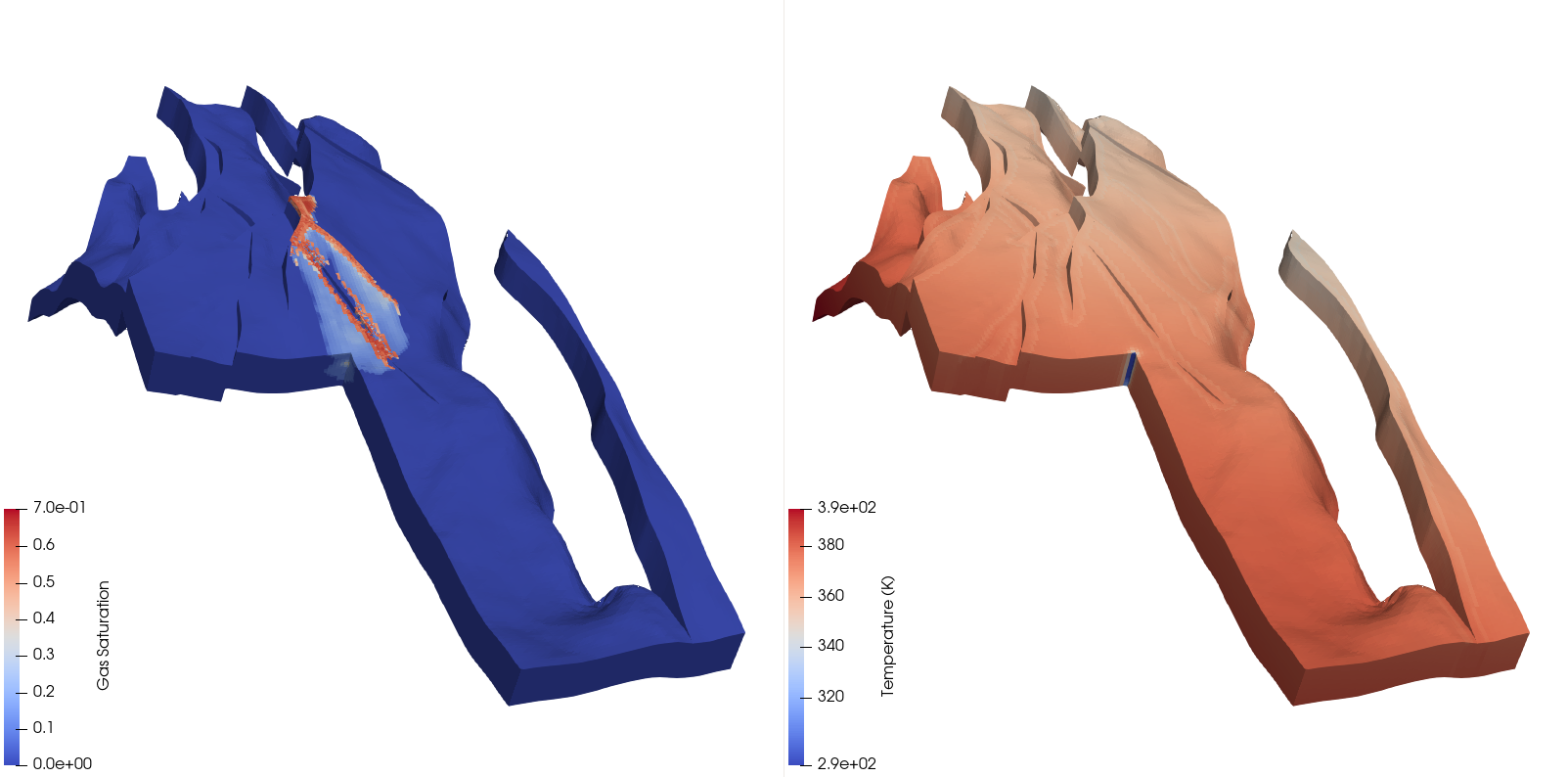}
    \caption{Gas saturation (left) and temperature (right) fields at the end of the simulation (225 years) for the Northern Lights case. Two half cuts are used to show the well line, and the Z direction is exaggerated by a factor of 10.}
    \label{fig:NL_end}
\end{figure*}

\subsection{High Island 24L}
\label{sec:hi24l}

The High Island 24L study area, located in the Texas waters of the Gulf of Mexico, is an approximately $12\times15$ km block highly suitable for CO$_2$ storage in its Miocene sand formation \cite{deangelo2019seismic,camargo2022deformation}. The geological model (and associated meshes) are based on seismic data and more than thirty wells drilled in the area, and contains 32 faults. The reservoir portion of the model is about 400-meter thick, with over- and under-burdens of 1.9 and 2.5 km, respectively. We have access to two different meshes for this model. The smaller mesh (denoted HI24L-S) has approximately 1.8M tetrahedral cells, and the larger mesh (denoted HI24L-L) approximately 7.8M tetrahedral cells. Both meshes conform to all sealed faults, and the reservoir part of the mesh is significantly higher resolution than the burdens, consisting of about 75\% of the total cells. Figure \ref{fig:HI24L_mesh} shows three planar views of the HI24L-S mesh, on which we can also observe that the near-fault areas are also more refined than the rest of the reservoir.

We assume that one well injects CO$_2$ roughly in the middle of the reservoir, in the bottom layer, at a rate of 1 Mt/year for 60 years. Figure \ref{fig:HI24L_end} shows the gas saturation and the temperature in the reservoir at the end of the simulation. Note that the temperature of the injected CO$_2$ is only 15$^\circ$C lower than the reservoir temperature, which is considered constant in this work. The plume moves upwards due to gravity effects, and since there is no migration period in this simulation, it is still expanding sideways.

\subsection{Northern Lights}
\label{sec:nlights}

The last test case studied in this work is part of the Northern Lights project, a joint venture between Equinor, Shell and TotalEnergies with plans to store large amounts of CO$_2$ in the Norwegian North Sea \cite{meneguolo2022impact,thompson2022characterization}. The storage site, named Aurora, is a saline aquifer located 100 km from the Norwegian coast, at a depth of 2,600 meters below the sea bed, and it is another (different from the first test case) part of the Johansen formation mentioned above. The model spans multiple kilometers squared, and the fully unstructured mesh, shown on Figure \ref{fig:NL_mesh}, has 3.5M cells and twenty fault lines, with the mesh conforming to all of them. Similar to the HI24L case, over and under burdens are included in the mesh to allow for geomechanics simulations to compute the long range pressure effects and the resulting deformations. Again, the reservoir is meshed with a much higher resolution than the burdens, and each fault area is also refined to be able to capture the flow patterns properly.

\begin{table*}[!htb]
    \centering
    \caption{Summary of the test cases properties. Each case has 4 Degrees of Freedom (DoFs) per cell. Injection is always done from a single well, roughly in the center and bottom of the grid.}
    \begin{tabular}{llrrrrrr}
    \toprule
               & Grid Type & Grid Size & DoFs & Well Rate & Injection & Migration & $\Delta$T ($^\circ$C) \\
    \midrule
    Johansen        & CPG & 58.3k & 0.23M & 1.00 Mt/yr & 25 yrs & 25 yrs & 90 \\
    Modified SPE10           & Cartesian & 660.0k & 2.64M & 0.01 Mt/yr & 116 days & $-$ & 68 \\
    HI24L-S & Unstruct. & 1,807.5k & 7.23M & 1.00 Mt/yr & 60 yrs & $-$ & 15 \\
    Northern Lights  & Unstruct. & 3,500.1k & 14.00M & 1.50 Mt/yr & 25 yrs & 200 yrs & $\sim$90 \\
    HI24L-L & Unstruct. & 7,752.2k & 31.01M & 1.00 Mt/yr & 60 yrs & $-$ & 15 \\
    \bottomrule
    \end{tabular}
    \label{tab:testCases}
\end{table*}

In the first 25 years of the simulation, one well injects pure CO$_2$ at a rate of 1.5 Mt/year from the bottom of the formation, then the well is shut down and the plume is allowed to migrate for 200 more years. Figure \ref{fig:NL_end} shows the gas saturation and the temperature profiles at the end of the migration period. We can see that the plume migrated upwards due to gravity, and has settled at the highest point of the connected reservoir section (all faults are assumed to be impermeable in the model).

\subsection{Summary}
\label{sec:summary}

Table \ref{tab:testCases} gives a summary of the different test cases, in order of increasing size. Modified SPE10 is an outlier in the operating conditions, with only a few months of injection and a much lower rate, due to the small size of the reservoir. Both Johansen and Northern Lights are full-cycle CCS cases, with both an injection period and a migration period, as well as a large temperature difference between the injected CO$_2$ and the reservoir temperature. HI24L cases do not have a migration period in this work, and the temperature difference is only 15$^\circ$C. Every case has four degrees of freedom (DoFs) per cell, so the size of the Jacobian matrix is four times the mesh size.

\section{Numerical Results}
\label{sec:results}

We use two different clusters to run the numerical experiments presented in this paper. The Sherlock cluster at Stanford has around 4,800 compute nodes split into four different kinds. The available nodes on the queue used in the work each have an Intel 5118 CPU with 24-core at 2.3 GHz, and 191 GB of RAM. Sherlock is only used for the smaller cases in this work because scheduling large jobs is quite challenging. For larger cases and all of the strong scaling results, the Quartz cluster at Lawrence Livermore National Laboratory is used. It has around 3,000 nodes, each of them with two Intel Xeon E5-2695 v4 processors, each with 18-core at 2.1 GHz and 128 GB of RAM.

Table \ref{tab:general} summarizes some relevant information about the five test cases. Importantly, both clusters have a 24 hour (1,440 min) maximum time for any computation. For some of the cases, they cannot be completed in that time and we have to adjust the analysis of the linear solver results to account for that. This is detailed in the next section.

\begin{table*}[!htb]
    \centering
    \caption{General information about the runs for all cases. Ranks refer to MPI ranks, and the simulations all have four degrees of freedom (DoF) per cell. Both the Sherlock and Quartz clusters have a maximum run time of 24 hours (1,440 min).}
    \begin{tabular}{lrrrrrrrr}
    \toprule
    & & & & & & & \multicolumn{2}{c}{Percentage}
    \\
    & & & & & \multicolumn{2}{c}{Elapsed Time} & \multicolumn{2}{c}{Completed}
    \\\cmidrule(lr){6-7}\cmidrule(lr){8-9}
               & Grid Size & Ranks & DoF/Rank & Cluster & CPR  & CPTR & CPR & CPTR \\
    \midrule
    Johansen        & 58.3k & 4 & 58.3k & Sherlock & 97 min & 8.7 min & 100.0\% & 100.0\% \\
    Modified SPE10           & 660.0k & 64 & 41.25k & Sherlock & 1,440 min & 161.8 min & 99.5\% & 100.0\% \\
    HI24L-S & 1,807.5k & 512 & 14.11k & Quartz & 1,440 min & 44.0 min & 55.3\% & 100.0\% \\
    Northern Lights  & 3,500.1k & 512 & 27.32k & Quartz & 1,440 min & 236.2 min & 10.2\% & 100.0\% \\
    HI24L-L & 7,752.2k & 1,024 & 30.28k & Quartz & 1,440 min & 247.4 min & 12.8\% & 100.0\% \\
    \bottomrule
    \end{tabular}
    \label{tab:general}
\end{table*}

\begin{figure*}[!htb]
    \centering
    \includegraphics[width=\textwidth]{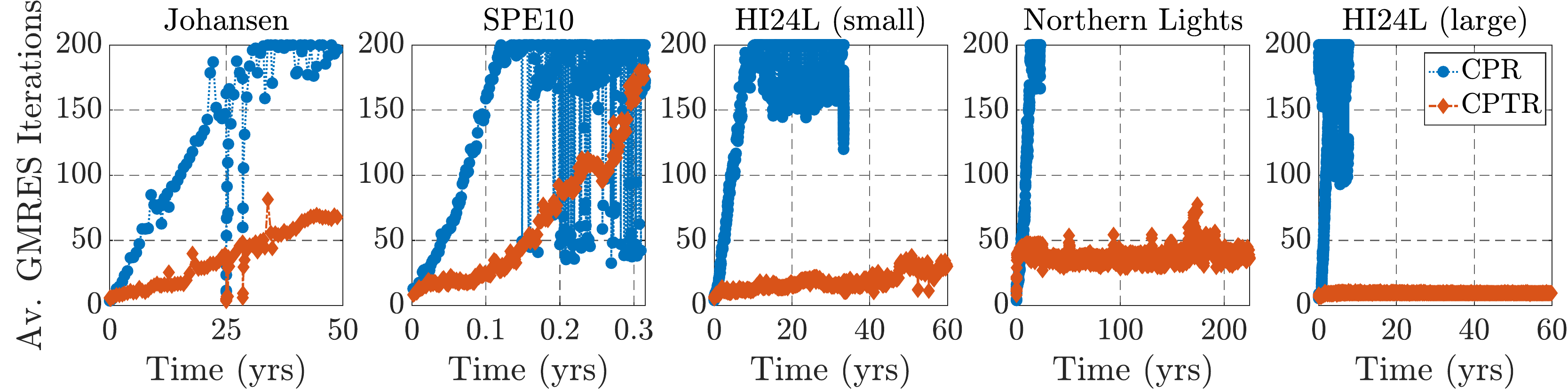}
    \caption{Average GMRES iterations vs physical time for various cases. Blue dotted line with circles is CPR, orange dashed line with diamonds is CPTR. From left to right, the test cases are Johansen, Modified SPE10, HI24L-S, Northern Lights, HI24L-L.}
    \label{fig:linit}
\end{figure*}

\subsection{Performance Analysis}
\label{sec:perf}

Our performance analysis is based on three main metrics:
\vspace{-0.2cm}
\begin{itemize} \itemsep 0pt
    \item[--] GMRES iterations (i.e. linear solver iterations),
    \item[--] GMRES solve time,
    \item[--] Preconditioner setup time.
\end{itemize}

The GMRES solve time depends on both the number of GMRES iterations and the cost of an individual iteration, which in turn depends on multiple factors, such as the cost of applying the preconditioner and the number of MPI ranks used in the simulation. Ideally, the iterations should never reach the maximum number allowed (in this work, 200), and stay as flat as possible during the simulation. Any other behavior suggests that the preconditioner is not providing a good enough approximation of the inverse, or that there are some issues with the test case (for example over-pressuring the reservoir) that makes it close to ill-posed, or that a very large time step specification is used. Note that time step considerations are important, but considered outside the scope of this paper and left for future work.

For the type of cases considered here, a good rule of thumb to qualitatively validate the performance of the preconditioner is to have the setup and solve times being roughly of the same order of magnitude. This is the expected behavior since physics-based block preconditioners are fairly expensive to set up, but should provide a high-quality approximate inverse and allow GMRES to converge in a small number of iterations \citep{Cremon20etal}. Another important point with GMRES is that the cost per iteration grows with the number of iterations, due to the increasing number of vectors that form the Krylov subspace basis.

Figure \ref{fig:linit} shows the average number of GMRES iterations per time step for all our test cases as a function of physical time. First, we see that for the three biggest and more realistic cases (HI24L-S, Northern Lights and HI24L-L), CPR never manages to finish the simulation in the 24 hours allowed on the cluster.
For all cases, CPR performs poorly and GMRES reaches the maximum iterations long before the end of the simulation, and stagnates from there. When that happens, GEOS accepts the solution at the last iteration and keeps trying to solve the system until the maximum number of nonlinear iterations, then cuts the time step. This results in both a large amount of wasted computations, and a much larger number of time steps required.

\begin{table*}[!htb]
    \centering
    \caption{Summary of the total iteration results over comparable physical time. Note that a GMRES failure occurs when the solver reaches its maximum number of iterations (200).}
    \begin{tabular}{lrrrrrrrrr}
    \toprule
    & & & \multicolumn{3}{c}{Total GMRES Iterations} & \multicolumn{2}{c}{GMRES Failures} & \multicolumn{2}{c}{Newton Iterations}
    \\\cmidrule(lr){4-6}\cmidrule(lr){7-8}\cmidrule(lr){9-10}
               & Grid Size & Ranks & CPR  & CPTR & Ratio & CPR  & CPTR & CPR  & CPTR\\
    \midrule
    Johansen & 58.3k & 4 & 96,546 & 16,043 & 6.0x & 701 & 0 & 687 & 496 \\
    Modified SPE10 & 660.0k & 64 & 798,072 & 65,530 & 12.2x & 7,505 & 0 & 5,245 & 1,047 \\
    HI24L-S & 1,807.5k & 512  & 1,586,541 & 34,118 & 46.5x & 17,132 & 0 & 8,677 & 2,342 \\
    Northern Lights & 3,500.1k & 512 & 1,439,553 & 44,353 & 32.5x & 9,512 & 0 & 7,577 & 1,040 \\
    HI24L-L & 7,752.2k & 1,024 & 906,629 & 25,446 & 35.6x & 10,205 & 0 & 5,422 & 2,803 \\
    \bottomrule
    \end{tabular}
    \label{tab:itResults}
\end{table*}

\begin{table*}[!htb]
    \centering
    \caption{Summary of the solve and setup time results (averaged per GMRES iteration). The setup time is the time taken to compute the preconditioner, and the solve time the time spent in the GMRES solver using the preconditioned system matrix.}
    \begin{tabular}{lrrrrrrrr}
    \toprule
    & & & \multicolumn{3}{c}{Average Setup Time (s)} & \multicolumn{3}{c}{Average Solve Time (s)}
    \\\cmidrule(lr){4-6}\cmidrule(lr){7-9}
               & Grid Size & Ranks & CPR  & CPTR & Ratio & CPR  & CPTR & Ratio \\
    \midrule
    Johansen        & 58.3k & 4 & 0.240 & 0.194 & 1.24x & 3.291 & 0.688 & 4.78x  \\
    Modified SPE10           & 660.0k & 64 & 0.602 & 0.642 & 0.94x & 12.101 & 4.023 & 3.01x  \\
    HI24L-S & 1,807.5k & 512 & 0.166 & 0.266 & 0.62x & 3.254 & 0.161 & 20.21x  \\
    Northern Lights  & 3,500.1k & 512 & 0.332 & 0.320 & 1.03x & 6.005 & 1.215 & 4.94x  \\
    HI24L-L & 7,752.2k & 1,024 & 0.284 & 0.406 & 0.70x & 5.127 & 0.172 & 29.81x  \\
    \bottomrule
    \end{tabular}
    \label{tab:timeResults}
\end{table*}

Going into more details about each case, Johansen uses the largest maximum time step of all test cases, around 200 days. Even though CPTR still does not require more than 80 GMRES iterations to converge, we do see an increase in the number of iterations with time increasing. On the Johansen and Modified SPE10 cases, CPTR greatly outperforms CPR, but it clearly starts to struggle towards the end of the simulation. We note here that reducing the time step would remove the iteration growth we observe on the Johansen and Modified SPE10 figures, and recover a flat curve. This suggests both that the maximum time step is too large, and more generally that the interaction between the linear and non-linear solvers can be improved. We deemed this out of the scope of this work, but will be an interesting consideration for practitioners observing a similar behavior.

The last three cases were the motivation for this work, as they are either actively considered for CCS projects in the North Sea (Northern Lights) or used as realistic analogs for future projects in the Gulf of Mexico (HI24L). Those meshes, properties and operating conditions are representative of the large-scale CCS cases planned in the near future. On those cases, CPTR performs admirably. It keeps the number of GMRES iterations well below 100, and is able to provide solutions when CPR can only get to a few tens of percent of the physical time. The temperature gradients are located in the near-well region, but they are enough to render the CPR preconditioner virtually inefficient. Even for the low-temperature gradient of the HI24L case (only 15$^\circ$C), CPR cannot handle the parabolic effects in the energy equation, whereas CPTR shows no issues and keeps the iteration numbers both low and constant. Northern Lights is the most challenging case, and CPTR is able to keep most of the time steps under 50 GMRES iterations.

Table \ref{tab:itResults} gives the detailed results of GMRES iterations, GMRES failures and Newton iterations. To keep the comparison fair, the table is only using the physical time that CPR was able to simulate. We note that CPTR does not show a single GMRES failure, and, as a result, shows a much lower number of Newton iterations. For the larger cases where most of the GMRES time is spent for discarded work, CPTR yields a 30-50x improvement in GMRES iterations. For Johansen and Modified SPE10, it shows a 6x and 12x improvement, respectively.

Table \ref{tab:timeResults} shows the average setup and solve times per GMRES iteration. The setup time is largely a function of the size of the problem and the implementation details within HYPRE \citep{Falgout02}. It is difficult to identify a clear trend, but using CPTR is within $\pm$35\% of the CPR cost. The solve time varies widely, mostly due to the amount of failures that CPR has to deal with. We observe anywhere from 3-30x improvements on the solve time alone.

\subsection{Strong Scaling}
\label{sec:strongScal}

\begin{figure*}[!htb]
    \centering
    \includegraphics[width=\textwidth]{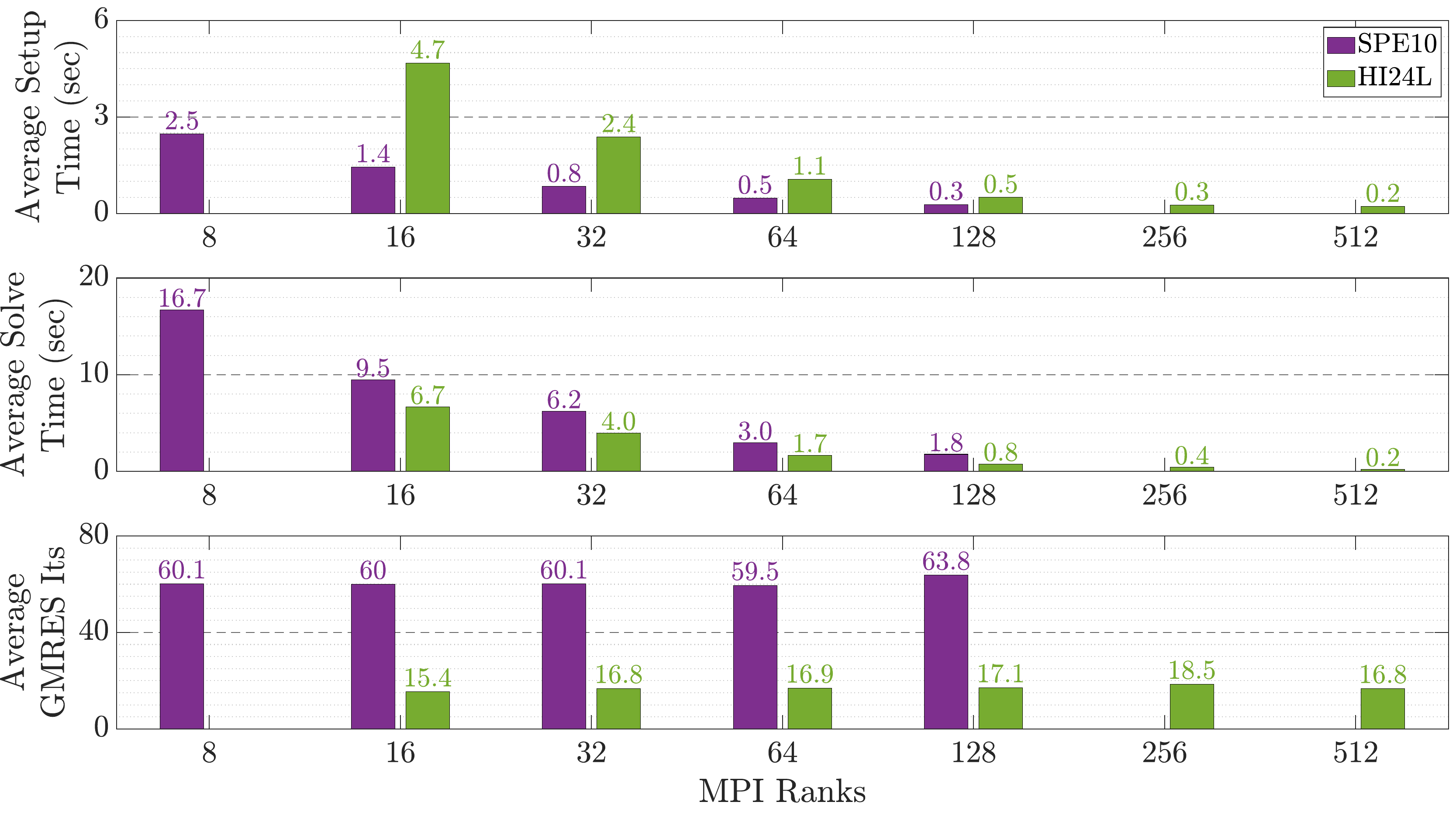}
    \caption{Average time to set up the preconditioner (top), average time to solve the linear system (middle) and average GMRES iterations (bottom) for the Modified SPE10 (purple) and the HI24L-S (green) cases, plotted as a function of the number of MPI ranks. Note that 8 cores would not run in 24 hours on Quartz for HI24L-S, and that 256/512 cores have too few DoFs per MPI rank for Modified SPE10, so they are omitted from the graphs. The reader is referred to the online version of the article for colored graphics.}
    \label{fig:strongScaling}
\end{figure*}

Strong scaling refers to the ability of an algorithm (and its parallel implementation) to provide faster solutions when given more resources, for a constant problem size. It is particularly  important for practitioners, especially compared to the more theoretical weak scaling. Typically, a given reservoir has a fixed size mesh (or maybe a few different resolutions), and the goal is to get results faster by adding more resources. Another issue with weak scaling is that it does not pair well with heterogeneous properties, nor with fully unstructured grids. For a given cell with associated properties, one would need a clever way to upscale or downscale the properties when changing the grid size, if one wants to solve the same physical problem. Although there are ways to go around those problems, due to the nature of this work and the test cases at hand, we selected strong scaling as the primary metric for parallel performance.

We use two of the previously described test cases in this section: Modified SPE10 and HI24L-S. All runs are done on the Quartz cluster, with a number of CPU cores, and by extension MPI ranks, ranging from 8 to 512. Some of those tests either take too long (e.g. HI24L-S with 8 cores) or have too few DoFs per rank (e.g. Modified SPE10 with 256 and 512 cores), so those runs are omitted. We seek to observe two things: a fairly constant number of iterations across MPI ranks, and a close-to-linear scaling of both the solve and setup times. Figure \ref{fig:strongScaling} shows the average results for both cases, in terms of GMRES iterations, solve time and setup time.

\begin{table*}[!htb]
    \centering
    \caption{Summary of the strong scaling results for both the Modified SPE10 and HI24L-S results. The HI24L case is a fully unstructured mesh, so the DoF/Rank is given as the average across all ranks.}
    \begin{tabular}{llrrrrrrr}
    \toprule
    & MPI Ranks & 8 & 16 & 32 & 64 & 128 & 256 & 512 \\
    \midrule
    \multirow{4}{*}{Modified SPE10} & DoF/Rank & 330k & 165k & 82.5k & 41.3k & 20.6k & -- & --  \\
    & Average GMRES Iterations & 54.61 & 54.49 & 54.57 & 54.02 & 54.91 & -- & --  \\
    & Average Setup Time (sec) & 2.47 & 1.44 & 0.84 & 0.47 & 0.28 & -- & -- \\
    & Average Solve Time (sec) & 16.71 & 9.47 & 6.21 & 2.95 & 1.77 & -- & -- \\
    \midrule
    \multirow{4}{*}{HI24L-S} & DoF/Rank & -- & $\sim$452k & $\sim$226k & $\sim$114k & $\sim$56k & $\sim$28k & $\sim$14k \\
    & Average GMRES Iterations & -- & 15.37 & 16.76 & 16.92 & 17.05 & 18.52 & 16.76 \\
    & Average Setup Time (sec) & -- & 4.67 & 2.38 & 1.06 & 0.50 & 0.26 & 0.21 \\
    & Average Solve Time (sec) & -- & 6.67 & 3.97 & 1.67 & 0.77 & 0.42 & 0.23 \\
    \bottomrule
    \end{tabular}
    \label{tab:strongScaling}
\end{table*}

\begin{figure*}[!htb]
    \centering
    \includegraphics[width=\textwidth]{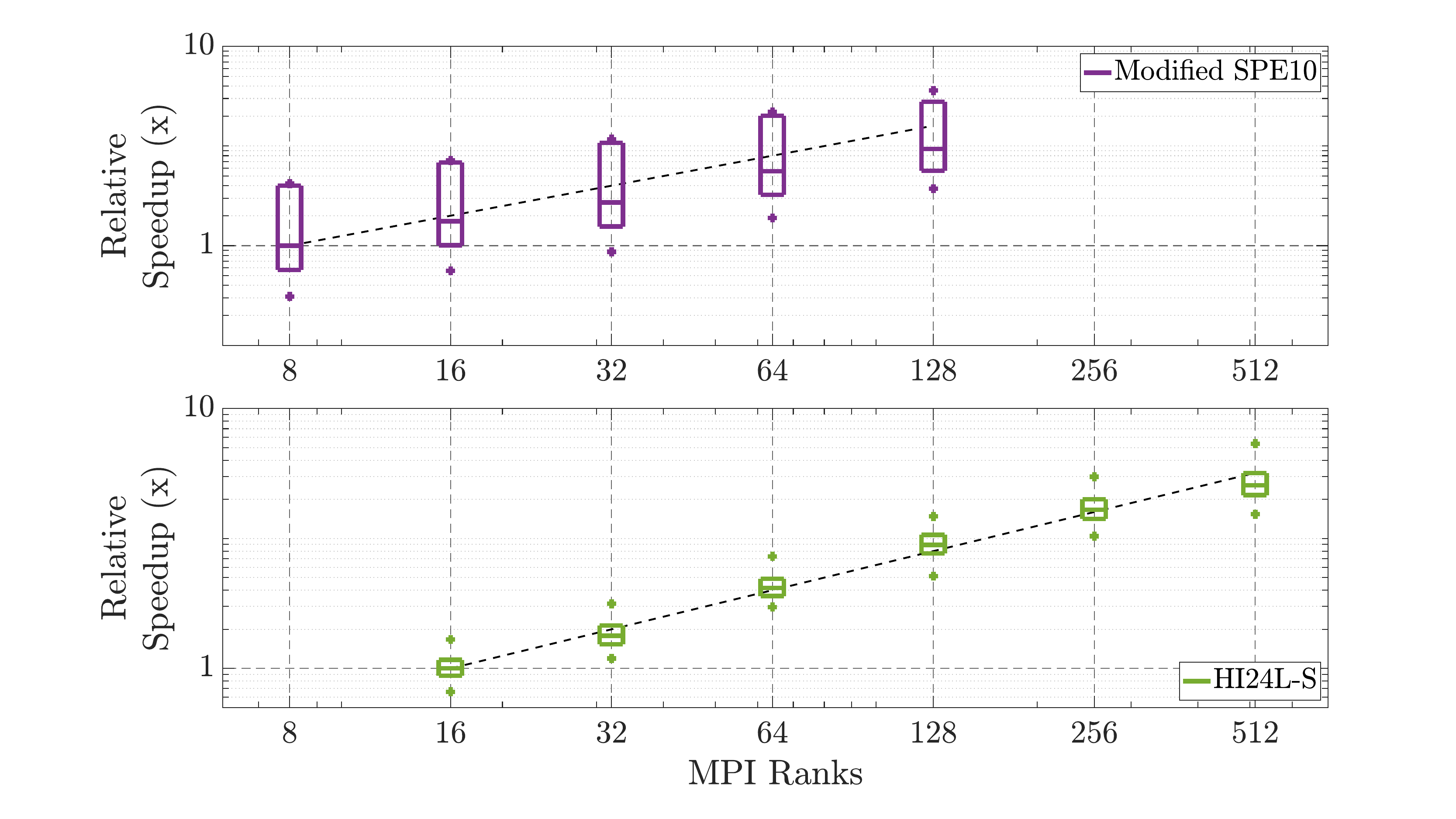}
    \caption{Relative speedup observed on the time spent in linear solver (setup + solve), for the Modified SPE10 and the HI24L-S case. The dashed line is the ideal linear scaling, based on the time taken by the 8-core and 16-core case, respectively. The line in each box is the average time spent in linear solve, and the boxes' top and bottom edges are plus/minus one standard deviation. The plus signs show the maximum/minimum value.}
    \label{fig:strongScalingBoxPlot}
\end{figure*}

In terms of GMRES iterations, the average for Modified SPE10 is 64,061 iterations, with a standard deviation of 3.2\%. For HI24L, the average is 16,573 iterations, with a standard deviation of 5.9\%. The solve time and setup time show a close-to-linear scaling, with the exception of the setup time for 512 cores on HI24L-S. This is likely due to the already low number of DoFs per rank, which does not allow enough work per rank to outweigh the communication costs. A detailed description of the results is given in Table \ref{tab:strongScaling}.

In terms of observed speedup, Figure \ref{fig:strongScalingBoxPlot} shows the Modified SPE10 and HI24L-S results in a box-plot manner and using a log-log scale. The speedup is virtually linear for HI24L-S, except the previously mentioned slight drop for 512 cores. For Modified SPE10, we do see a good speedup (about 60\% of the ideal 8-core based scaling for the 128-core case), but the challenges of the model do not allow it to be linear. The nature of the problem, being highly heterogeneous, and the large time steps are very challenging for the load balancing, and we can observe that with the large standard deviation in the box plot. In general, using more cores than around 20k degrees of freedom per core will systematically result in poor performance. In those cases, the time required to setup the preconditioner will start to dominate, since there is not enough compute to do in the solve step. This behavior has been observed with HYPRE in other cases as well \citep{Falgout02, Baker11}.

\begin{figure*}[!htb]
    \centering
    \includegraphics[width=\textwidth]{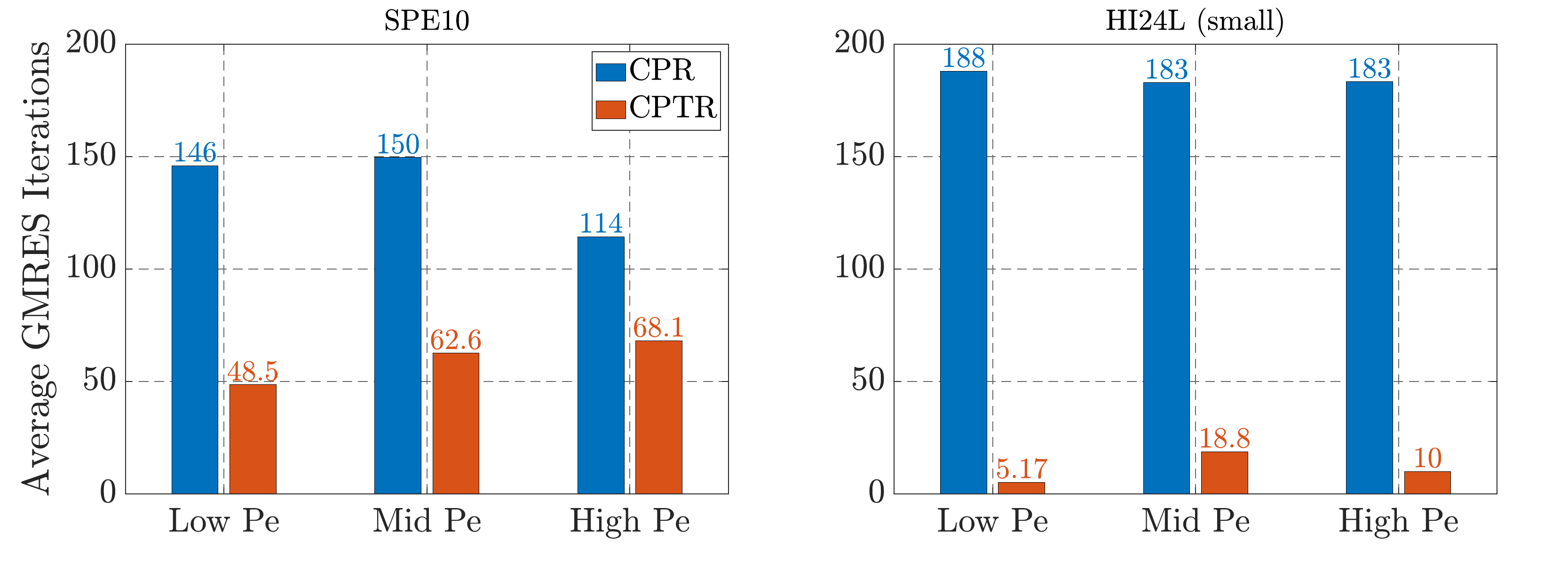}
    \caption{Performance of CPR and CPTR for a range of thermal Pe numbers, on the Modified SPE10 case (left) and the small HI24L case (right).}
    \label{fig:linit_Pe}
\end{figure*}

Overall, the strong scaling performance of CPTR in cases with no time-step issues is virtually linear, and in line with the previously observed performance of GEOS (using HYPRE's MGR and BoomerAMG) for physics-based preconditioners. This work is the first to use the system version of BoomerAMG, paired with a two-level MGR with Quasi-IMPES decompositions. Although these building blocks have been tested separately \citep{Baker11, Bui20, Bui21}, it is valuable to confirm that the scaling does not suffer from the increased complexity and coupling of the global system and all components are performing as expected.

\subsection{Sensitivity to Thermal P\'eclet Number}
\label{sec:peclet}

When running cases involving the energy equation, a dimensionless number arises to quantify the relative importance of conduction and convection. The thermal P\'eclet number, Pe, is defined as
\begin{equation} \label{eq::pe}
    \textrm{Pe} = \dfrac{q\rho c_p}{\kappa L} \,,
\end{equation}
with $L$ a characteristic length, $q$ a characteristic flow rate, $\rho$ and $c_p$  the density and the specific heat capacity of the fluid carrier and $\kappa$ the global thermal conductivity. Pe is infinite when conduction is negligible ($\kappa = 0$), and zero when convection is negligible  ($c_p = 0$). Importantly, since the energy equation is parabolic, those Pe asymptotes correspond to the pure hyperbolic or pure elliptic cases, respectively. Recall that for the CPR preconditioner, the energy equation is treated as a secondary variable. If the equation is sufficiently hyperbolic, CPR should be able to perform better. Note that due to the presence of numerical diffusion, we never reach a pure hyperbolic case in practice. We identified in \citet{Cremon20etal} that the CPTR preconditioner drastically reduced the sensitivity to Pe, and was able to perform efficiently on thermal oil displacement problems, with and without chemical reactions. This section aims to confirm those results for thermal CCS cases.

Similarly to the previous section, we use the Modified SPE10 and HI24L-S cases to investigate the performance of the preconditioner for thermal CCS cases. Figure \ref{fig:linit_Pe} shows the results for three different values of the Pe number. The Mid case corresponds to the default values for the reservoirs considered here. High and Low cases decrease (resp. increase) the value of the overall conduction ($\kappa$) by two orders of magnitude.

On the Modified SPE10 case, CPR does perform better for the High Pe number case, by around 23\%. However, it still never achieves fewer than 114 iterations (on average) and is outperformed by CPTR in every case, by 40-67\%. On the HI24L case, like we previously observed, CPR is not able to perform at all and virtually always fails (going to 200 iterations). This is due to the much larger size of the reservoir, leading to very few cells experiencing fronts over the course of the simulation. The long-range elliptic-like components of the temperature equation are always dominant in the solver, and CPR can never perform well. CPTR is able to stay below 20 average iterations in all cases. These examples cover the typical range of values encountered in thermal CCS cases for saline aquifers and confirm that CPTR is performing admirably on all cases.





\section{Conclusion \& Discussion}
\label{sec:conclusion}

In this work, we studied the performance of the CPTR preconditioner, introduced in \citet{Roy20} and \citet{Cremon20etal}, on a number of different test cases for large-scale thermal CO$_2$ injection cases. We used a standard GMRES iterative solver and compared the performance of CPTR with the state-of-the-art CPR preconditioner, which has been known to perform poorly on thermal cases.

We showed that CPTR results in orders of magnitude less GMRES iterations and required linear solver time for multiple realistic CCS cases. The test cases include both Cartesian and fully unstructured meshes, up to tens of millions of degrees of freedom. In all cases CPTR is able to compute the solution in a few hours using tens or hundreds of CPU (depending on the case size), where CPR would routinely reach the 24-hour limit of the two clusters used to run the simulations. In addition, we confirmed that the parallel scaling of the implementation in GEOS is close to linear, and that the preconditioner is insensitive to the thermal P\'eclet number (which compares thermal diffusion to thermal advection). This ensures that for a variety of physical and numerical conditions, CPTR will perform optimally on thermal CCS cases.

As future work, it would be interesting to confirm our results for more complex physics, for example more complex mixtures representing not only saline aquifers but also depleted hydrocarbon reservoirs.In this work, we did not vary the parameters of the AMG and ILU routines, but sensitivity studies could yield valuable information about the behavior of the preconditioner. We stated in a previous section that time stepping strategies can have a large effect on the simulation time. Devising a scheme to a priori determine a well-performing scheme using CPTR would help practitioners a lot. Finally, everything in this work is done using CPU clusters, studying the performance on other types of hardware accelerators would confirm that the algorithm can always perform well.

\section*{Acknowledgements}
\label{sec:ackn}
We acknowledge the Livermore Computing Center and Stanford University for providing the computational resources used in this work.
We thank Nicola Castelletto for his insights and fruitful discussions, as well as Quan M. Bui, Daniel Osei-Kuffuor and Victor A. Paludetto Magri for the many MGR developments used in the present article.
We thank Bernd Flemisch and Holger Class for providing access to the Johansen dataset, and the Northern Lights Joint Venture for providing access to the Northern Lights dataset.

\section*{Funding and Competing Interests}

Funding for FPH and JF was provided by TotalEnergies S.E. through the FC-MAELSTROM project.

The authors have no competing interests to declare that are relevant to the content of this article.



\bibliographystyle{elsarticle-num-names}
\bibliography{bib_final}





\end{document}